\documentclass[
reprint,
superscriptaddress,
amsmath,amssymb,
aps,
prl,
]{revtex4-1}

\usepackage[utf8]{inputenc}
\usepackage{amsfonts}
\usepackage{hyperref}
\usepackage{graphicx}
\usepackage{mathptmx}
\usepackage{textcomp}

\usepackage{siunitx}
\DeclareSIUnit\molar{\mole\per\cubic\deci\metre}
\DeclareSIUnit\Molar{M}

\graphicspath{{Figures/}}

\begin{document}

\title{Dynamic bidirectional coupling of membrane morphology and rod organization in flexible vesicles}

\author{Stijn van der Ham}
    \thanks{These authors contributed equally to this work.}
    \affiliation{Active Soft Matter and Bio-inspired Materials Lab, Faculty of Science and Technology, MESA+ Institute, University of Twente, 7500 AE Enschede, The Netherlands}
\author{Andr\'e F. V. Matias}
    \thanks{These authors contributed equally to this work.}
    \affiliation{Soft Condensed Matter \& Biophysics, Debye Institute for Nanomaterials Science, Utrecht University, Princetonplein 1, 3584 CC Utrecht, The Netherlands}
\author{Marjolein Dijkstra}
    \email{m.dijkstra@uu.nl}
    \affiliation{Soft Condensed Matter \& Biophysics, Debye Institute for Nanomaterials Science, Utrecht University, Princetonplein 1, 3584 CC Utrecht, The Netherlands}
\author{Hanumantha Rao Vutukuri}
    \email{h.r.vutukuri@utwente.nl}
    \affiliation{Active Soft Matter and Bio-inspired Materials Lab, Faculty of Science and Technology, MESA+ Institute, University of Twente, 7500 AE Enschede, The Netherlands}

\begin{abstract}
    The ordering of rod-like particles in soft, deformable containers emerges from the interplay of anisotropic interactions, geometric confinement, and boundary compliance. This competition couples internal particle organization to container morphology, producing behavior distinct from both rigid confinement and bulk systems. Such coupling is also relevant to biological contexts in which filamentous structures are confined by deformable membranes. Using a minimal model combining experiments and simulations of colloidal rods encapsulated in lipid vesicles, we show that soft confinement drives a bidirectional coupling between internal order and vesicle shape. This interplay gives rise to a phase diagram in which elongated vesicles promote nematic alignment at lower packing fractions, whereas higher packing fractions induce smectic-like ordering that reshapes vesicles into plate-like morphologies with increased bending energy. Furthermore, by controlling vesicle volume and membrane area, we demonstrate that boundary conditions enable reversible tuning of both vesicle shape and internal rod organization. These results establish a framework for dynamically controlling colloidal self-assembly in soft containers and provide insight into the organization of anisotropic building blocks in deformable, cell-like, confinements.
\end{abstract}

\maketitle

Liquid crystals composed of shape-anisotropic particles dispersed in a fluid are a longstanding topic of interest in soft matter physics~\cite{de1993physics}. In bulk suspensions, rod-like colloidal particles exhibit phase behavior that is primarily governed by packing fraction and aspect ratio, giving rise to isotropic, nematic, and smectic phases~\cite{onsager1949effects,frenkel1988thermodynamic,Bolhuis1997}. When these particles are confined to finite volumes, the imposed geometric and topological constraints can dramatically modify anchoring conditions, defect formation, and intrinsic ordering~\cite{leferink2013phase,Jull2024,camposvillalobos2025shapingboundariescontroltransport}. Such confined systems are commonly analyzed through the topology and structure of the defects that emerge. Extensive studies of colloidal nematic and smectic phases in rigid, quasi-two-dimensional geometries, including circles, annuli, and polygonal containers, have revealed clear correlations between boundary geometry, defect configurations, and emergent order~\cite{wittmann2021particle,cortes2016colloidal,monderkamp2021topology,garlea2015defect,garlea2016finite}.

In contrast to rigid confinement, far less is understood about how rod-like particles organize within deformable boundaries. Flexible containers can deform in response to internal stresses, partially accommodating particle packing and thereby altering emergent structures. Similar effects have been observed for finite numbers of spheres confined within deformable containers, where boundary flexibility favors elongated, “sausage-like” arrangements over compact clusters, in contrast to rigid confinement and bulk behavior~\cite{MarinAguilar2023sausage}.

Soft confinement of anisotropic or filamentous objects is also common in biological systems, where cytoskeletal filaments, filamentous viruses, or rod-shaped bacteria are enclosed by flexible membranes~\cite{fletcher2010cell,isberg2009legionella,kolehmainen1965structure}. Under strong confinement, these filaments experience spatial constraints that influence their orientational order~\cite{alvarado2014alignment,e2011self,Mulder2010,Lindeboom2013,Hawkins2010}, while stresses generated by filament assembly may deform the surrounding membrane~\cite{sens2015membrane,fletcher2010cell,pollard2009actin}. Similar coupling has been observed in encapsulation experiments with giant unilamellar vesicles (GUVs)~\cite{dimova2019giant}, where actin filaments or microtubules reorganize and reshape vesicle morphologies~\cite{shi2023morphological,emsellem1998vesicle,tsai2015shape,tanaka2018repetitive,agudo2025synthetic,sciortino2025active}. Although these systems involve additional complexities such as filament flexibility, activity, and biochemical regulation, they highlight the importance of generic physical coupling between anisotropic objects and deformable boundaries.

Despite the fundamental interest in soft confinement of rod-like particles, a systematic understanding of their packing within deformable containers remains elusive. Here, we investigate a minimal model consisting of rigid colloidal rods encapsulated inside flexible GUVs, allowing us to isolate the interplay between anisotropic particle ordering and membrane deformability in a controlled and tunable setting.

Combining experiments and simulations, we construct a phase diagram of rod organization and vesicle morphology as functions of vesicle volume and rod packing fraction. At low packing fractions, elongated vesicle shapes promote nematic alignment, effectively shifting the isotropic–nematic transition to lower densities. At higher packing fractions, frustration of the smectic order generates mechanical stresses that reshape vesicles into plate-like morphologies, accompanied by an increase in membrane bending energy. In this regime, the resulting smectic layers give rise to faceted vesicle shapes, with curvature localized near rod tips and suppressed along the rod bodies. Together, these results reveal a strong, bidirectional coupling between rod organization and membrane deformability. Finally, by modulating vesicle volume or membrane area, we induce reversible and controlled transitions between isotropic, nematic, and smectic phases, as well as between elongated and plate-like vesicle morphologies, highlighting the versatility of GUVs as soft, responsive containers.

\section*{Results and discussion}

\subsection*{Experimental and numerical realization of rods encapsulated in vesicles}

\begin{figure}[t!]
	\centering
	\includegraphics[width=0.95\linewidth]{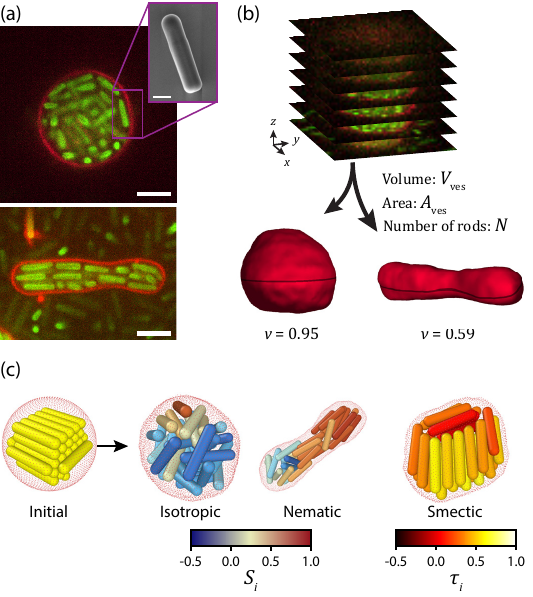}
	\caption{Experimental and numerical model setup. (a) Composite confocal fluorescence images of the midplane of a spherical isotropic vesicle and a linear nematic vesicle. The vesicle membrane is shown in red, and the encapsulated silica rods in green. The inset shows a SEM image of a representative silica rod. Scale bars: \qty{5}{\um} (confocal) and \qty{1}{\um} (inset). (b) 3D reconstruction of vesicles from confocal fluorescence $z$-stacks. Image stacks were segmented using the LimeSeg plugin in FIJI (ImageJ)~\cite{machado2019limeseg,schindelin2012fiji}, enabling quantitative measurements of vesicle volume $V_\mathrm{ves}$, surface area $A_\mathrm{ves}$, and number of rods $N$. (c) Coarse-grained simulations in LAMMPS, consisting of rigid rods encapsulated within a meshless membrane. Isotropic, nematic, and smectic rod ordering is obtained by gradually decreasing the vesicle volume from an initially spherical, isotropic configuration. The rod color corresponds to their local nematic order $S_i$ (isotropic and nematic vesicles), and local smectic order $\tau_i$ (smectic vesicle).}
	\label{fig:setup}
\end{figure}

To investigate the orientational order of colloidal rods confined within a deformable container, we encapsulated fluorescent silica rods inside GUVs (see Fig.~\ref{fig:setup}a and Supporting Movie~{S1}) using the droplet-transfer method~\cite{vutukuri2020active} (Materials and Methods). The vesicle membranes were labeled with Liss Rhod PE, while the rods were coated with a fluorescein isothiocyanate (FITC)-containing silica layer, enabling two-color confocal fluorescence imaging and three-dimensional reconstruction of both the vesicle boundary and its interior. The main population of rods had an average length $L = \qty{4.5}{\um}$ and diameter $D = \qty{1}{\um}$ (see Supporting Fig.~{S1}), yielding an aspect ratio of $L/D = 4.5$. Vesicle diameters ranged from 6 to \qty{14}{\um}, encapsulating between 10 to 100 rods per vesicle. This regime, where rod and vesicle length scales are comparable, is expected to produce the strongest coupling between internal ordering and membrane shape.

To complement the experiments, we replicate the experimental conditions \textit{in silico}. The system was modeled using the Large-scale Atomic/Molecular Massively Parallel Simulator (LAMMPS)~\cite{Thompson2022}, where each rod was represented as a rigid body composed of five non-overlapping spheres, yielding an aspect ratio of $L/D = 5$. The vesicle was modeled by a meshless membrane as described in the Methods and in Refs.~\cite{Yuan2010,Fu2017}. Outside the vesicle, we mimic the osmotic pressure in the experiments by simulating an explicit solvent consisting of spherical particles in the \textit{NPT} ensemble. This sets the vesicle volume. The surface area of the vesicle was controlled by the number of membrane particles. With the exception of vesicle-vesicle interactions, pairwise interactions were described by the Weeks-Chandler-Andersen (WCA) potential (see Methods). Simulations were initialized with ordered rods inside a spherical vesicle, as seen in Fig.~\ref{fig:setup}c. To reach equilibrium, we slowly increased the pressure of the explicit solvent to the target value (Supporting Movie~{S2}), after which we took time averages of all the relevant quantities.

To quantify the coupling between vesicle shape and rod packing, we measured two key parameters: the packing fraction of the encapsulated rods and the vesicle's reduced volume, a dimensionless measure of vesicle shape \cite{seifert1991shape}. The reduced volume, $\nu$, is defined as the ratio of the vesicle volume, $V_\mathrm{ves}$, to the volume of a sphere, $V_{\mathrm{sph}}$, with the same surface area, $A_\mathrm{ves}$,
\begin{equation}
	\nu = \frac{V_\mathrm{ves}}{V_{\mathrm{sph}}} = 3\sqrt{4\pi} \frac{V_\mathrm{ves}}{A_\mathrm{ves}^{3/2}} \\.
	\label{eq:reduced_volume}
\end{equation}
Here, $\nu = 1$ corresponds to a perfect sphere, while decreasing values indicate increasing deviations from sphericity (e.g. ellipsoidal or deflated morphologies) associated with low membrane tension; in the limiting case of an infinitely long line or plane $\nu \to 0$~\cite{Jari1995}. Experimentally, $\nu$ was determined by extracting $V_\mathrm{ves}$ and $A_\mathrm{ves}$ from 3D confocal reconstructions using a contour-based segmentation algorithm (Fig.~\ref{fig:setup}b)~\cite{machado2019limeseg,van2025shallow}. In simulations, these quantities were extracted directly from the meshless membrane representation (see Methods).

The rod packing fraction, $\eta$, was calculated from the same confocal image data by counting the total number of rods in the vesicle interior, $N$, multiplying by the average rod volume, and dividing by the vesicle volume, such that
\begin{equation}
	\eta = \frac{N V_\mathrm{rod}}{V_\mathrm{ves}} \\.
	\label{eq:packing_fraction}
\end{equation}
For the experiments, the average rod volume is $V_\mathrm{rod}=\qty{3.1}{\um^3}$ (see Methods and Supporting Fig.~{S1}), while in simulations it corresponds to the volume of five spheres, $5\pi\sigma^3/6$. We note that, although this defines the simulated rod volume, it slightly underestimates the effective packing fraction compared to fully solid rods.

\begin{figure*}[t!]
	\centering
	\includegraphics[width=1\linewidth]{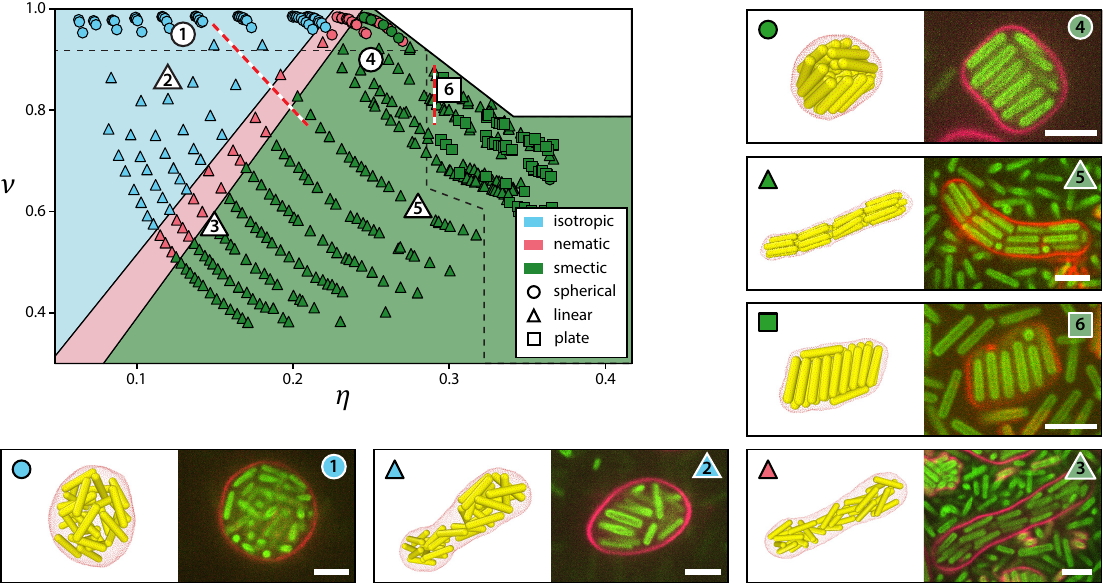}
	\caption{Phase diagram of rod order and vesicle shape as a function of packing fraction $\eta$ and reduced volume $\nu$, for vesicles encapsulating $N = 30$ rods with aspect ratio $L/D = 5$. Colored data points indicate simulation results, while numbered white markers correspond to representative experimental vesicles. The colors indicate the rod ordering: blue for isotropic, red for nematic, and green for smectic. The symbols denote the vesicle shape: circles for spherical, triangles for linear, and squares for plate-like vesicles. The background colors denote the approximate boundaries of the different rod ordering regimes, while the black dashed lines distinguish the spherical, linear, and linear/plate regions. Representative snapshots from both experiments and simulations are shown for different phases, corresponding to the numbered markers. The white region in the top-right corner is inaccessible, as spherical vesicles (high $\nu$) are geometrically incompatible with very high rod packing. The red-white dashed lines correspond to the volume and area modulation explored in Fig.~\ref{fig:VolumeChange}. Scale bars: \qty{5}{\um}.}
	\label{fig:phase}
\end{figure*}

Additionally, for both the experiments and simulations, we classify the nematic order of the rods using the average local nematic, $\langle S_i\rangle$~\cite{Cuetos2007}, and smectic, $\langle\tau_i\rangle$~\cite{Bakker2016}, order parameters (see Methods). Low $\langle S_i\rangle$ and $\langle\tau_i\rangle$ corresponds to isotropic ordering, high $\langle S_i\rangle$ and low $\langle\tau_i\rangle$ corresponds to nematic ordering, and high $\langle S_i\rangle$ and $\langle\tau_i\rangle$ corresponds to smectic ordering. Representative simulation snapshots at different $\eta$ and $\nu$ values are shown in Fig.~\ref{fig:setup}c, illustrating transitions from isotropic to nematic and smectic-like ordering as the packing fraction increases, evidenced by the increase in the local order parameters. 

The vesicles exhibited a wide variety of shapes, giving rise to diverse rod arrangements. At low rod packing fractions, rods were predominantly isotropic, while increasing the packing fraction promoted nematic and eventually smectic-like ordering, in agreement with bulk phase behavior~\cite{Bolhuis1997,Dussi2018}. However, unlike in bulk, confinement within deformable vesicles introduced an additional coupling between rod organization and vesicle shape. Vesicles frequently deformed to accommodate local rod alignment, while, conversely, rod orientation adapted to the evolving membrane geometry.

\subsection*{Phase diagram}

To investigate how vesicle shape couples to the packing order of confined rods, we constructed a phase diagram of the reduced volume $\nu$ and packing fraction $\eta$ (Fig.~\ref{fig:phase}), integrating simulation and experimental data. In simulations, vesicles evolved along trajectories of approximately constant membrane area while their internal volume decreased for a fixed number of rods ($N=30$) (Supporting Fig.~{S2}). Experimental vesicles contained a comparable number of rods on average ($\langle N \rangle \approx 31$). When experimental configurations are mapped onto the simulated phase diagram, we find good quantitative agreement between simulations and experiments in both vesicle morphology and internal rod organization (Table~{S1}).

Systematic variation of $\eta$ and $\nu$ in simulations reproduces the full range of experimentally observed morphologies and reveals distinct packing regimes, namely isotropic, nematic, and smectic, identified using the order parameters defined in Methods. Representative experimental configurations are shown in snapshots 1–6. Snapshots 1 and 2 correspond to isotropic packing, whereas snapshots 4–6 exhibit smectic order. Snapshot 3 lies close to the nematic–smectic boundary and is classified as smectic according to the applied criteria, although its layering is less pronounced than in deeper smectic states (e.g., snapshot 5).

To further quantify the observed configurations, each configuration was classified into one of three morphological categories, spherical (snapshots 1 and 4), linear (snapshots 2, 3, and 5), or plate-like (snapshot 6), based on the number of symmetry axes, as quantified by the asphericity and acylindricity (see Methods). Applying the same shape metrics to experimental vesicles yielded consistent classifications between experiment and simulation.

The resulting phase diagram demonstrates that the packing of colloidal rods within flexible vesicles is governed not only by the packing fraction $\eta$, but also by vesicle shape, quantified by the reduced volume $\nu$ (Fig.~\ref{fig:phase})~\cite{seifert1991shape}. This behavior contrasts sharply with bulk rod suspensions, where ordering is controlled solely by $\eta$~\cite{Dussi2018, Bolhuis1997}. Under confinement, decreasing $\nu$ produces increasingly elongated vesicles, which promote rod alignment along the vesicle’s long axis~\cite{seifert1991shape}. This geometry-induced alignment stabilizes nematic order and facilitates the formation of smectic layers oriented along the same axis (snapshots 3 and 5). Nevertheless, packing fraction remains essential: at sufficiently low $\eta$, even strongly elongated vesicles remain isotropic (snapshot 2). Overall, the combined effect of $\eta$ and $\nu$ produces a pronounced shift of the isotropic–nematic transition to lower packing fractions as vesicles become more elongated, a trend that persists across different system sizes and rod aspect ratios (Supporting Fig.~{S3}). In particular, increasing the aspect ratio shifts the transitions to even lower packing fractions and widens the nematic region, akin to bulk systems~\cite{Bolhuis1997, Dussi2018}.

\begin{figure}[t!]
	\centering
	\includegraphics[width=1.0\linewidth]{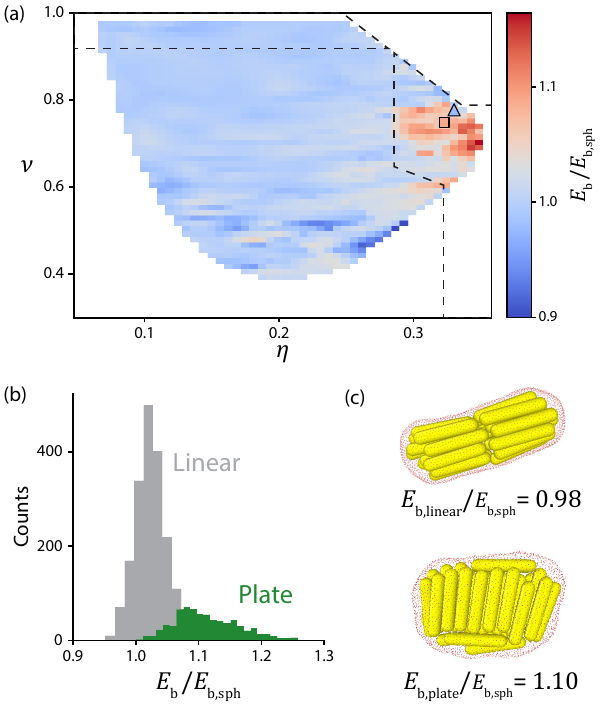}
	\caption{Bending energy of the vesicle ($E_\mathrm{b}$) rescaled by the bending energy of a vesicle containing spheres ($E_\mathrm{b,sph}$). (a) Color map of the rescaled bending energy as a function of the vesicle's reduced volume $\nu$ and rod packing fraction $\eta$. Blue (red) color indicates low (high) rescaled bending energy. The black dashed lines correspond to different vesicle shape regions (Fig.~\ref{fig:phase}) The red area corresponds to the coexistence region of linear and plate-like vesicles. Highlighted are the linear (triangle) and plate-like (square) morphologies from (c). (b) Histogram of the rescaled bending energy within the coexistence region, the gray (green) histogram corresponds to linear (plate-like) vesicles. (c) Representative simulation snapshots of a linear and plate-like vesicle with their corresponding bending energy annotated.}
	\label{fig:bending_energy}
\end{figure}

\subsection*{Rods reshape vesicle morphology}
Having established that vesicle shape influences rod ordering, we next demonstrate that the reverse is also true: rods actively reshape the vesicle. This is supported by the emergence of plate-like vesicle morphologies, which appear only at high packing fractions and coexist with linear vesicles (see Fig.~\ref{fig:phase}). Notably, these plate-like shapes occur exclusively in vesicles containing rods, whereas vesicles encapsulating spherical particles do not adopt such morphologies (Supporting Fig.~{S4}). Their formation thus reflects active deformation by smectic rod packing, driving vesicles away from the minimum-energy shapes dictated solely by surface tension and osmotic pressure.

This effect is quantified by comparing the bending energy of rod-containing vesicles, $E_\mathrm{b}$, with that of vesicles containing spheres, $E_\mathrm{b,sph}$, at identical $\eta$ and $\nu$ (Materials and Methods). The resulting ratio, $E_\mathrm{b}/E_\mathrm{b,sph}$, shown in Fig.~\ref{fig:bending_energy}a, remains close to unity across most of the parameter space but increases sharply in the plate–linear coexistence region at high $\eta$, indicating that vesicles undergo energetically costly deformations in this regime. To verify that this increase is specifically associated with plate-like shapes, Fig.~\ref{fig:bending_energy}b compares the bending-energy ratios $E_\mathrm{b}/E_\mathrm{b,sph}$ for linear and plate-like vesicles within the coexistence region, illustrating that plate-like vesicles exhibit systematically higher bending energies (see also representative examples in Fig.~\ref{fig:bending_energy}c). Their lower abundance relative to linear vesicles is therefore consistent with the higher energetic cost required to maintain the plate-like morphology. We remark that, despite the higher energy cost, plate-like vesicles remain present when the compression rate from spherical to plate-like shapes is decreased by an order of magnitude.

\subsection*{Coupling between membrane curvature and rod orientation}

To understand the mechanism underlying these global shape transformations, we next examine the local coupling between rod orientation and membrane curvature. In particular, we quantify how curvature varies along the vesicle surface and how these variations correlate with rod position and anchoring orientation.

\begin{figure}[t!]
	\centering
	\includegraphics[width=1.0\linewidth]{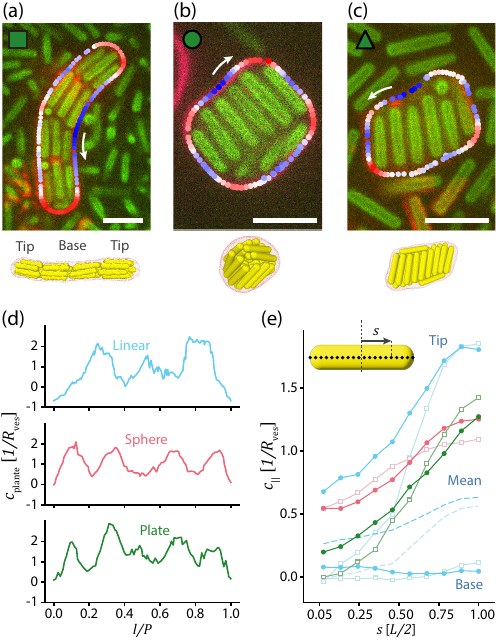}
	\caption{Correlation between membrane curvature and rod ordering in smectic vesicles with different morphologies.
		(a–c) Confocal midplane cross-sections overlaid with the in-plane membrane curvature $c_\mathrm{plane}$, defined as the curvature of the two-dimensional vesicle contour in the imaging plane (red: high curvature; blue: low curvature), given for a smectic linear (a), sphere (b), and plate-like (c) morphology. Below each image, a representative simulation snapshot of a corresponding vesicle morphologies is shown. Scale bars: \qty{5}{\um}.
		(d) In-plane membrane curvature $c_\mathrm{plane}$, expressed in units of $1/R_\mathrm{ves}$, where $R_\mathrm{ves} = (3V_\mathrm{ves}/4\pi)^{1/3}$ is the vesicle radius, plotted as a function of the normalized arc length $l/P$ along the vesicle outline. Here, $l$ is the curvilinear distance measured along the two-dimensional vesicle contour and $P$ is the total contour perimeter. 
		(e) Correlation between the position along the rod, $s$ (schematic inset), expressed in units of $L/2$, and the membrane curvature along the rod axis, $c_{\parallel}$ (see Methods), for linear (cyan), spherical (red), and plate-like (green) morphologies. Data represent averages over all rod segments within \qty{1.5}{\um} of the membrane. Experimental and simulation results are shown by filled circles and open squares, respectively. For linear vesicles, rods near the vesicle tips and along the base exhibit distinct curvature responses; their mean behavior is indicated by the dashed line.}
	\label{fig:curv_corr}
\end{figure}

Linear vesicles display a clear anchoring pattern: rods in the high-curvature tip regions exhibit homeotropic alignment, whereas rods in the low-curvature midsection adopt planar anchoring (Fig.~\ref{fig:curv_corr}a,d). This behavior parallels curvature-dependent anchoring reported for smectic rods confined in rigid 2D elliptical wells, where planar alignment persists up to a critical wall curvature $c^* \approx 1/(1.6L)$~\cite{Jull2024}. Motivated by this connection, we first evaluated whether an analogous curvature threshold could explain the anchoring transition in 3D vesicles. Because membrane curvature is directional in three dimensions, we computed the curvature projected along the rod axis (Supporting Fig.~{S5}). For large vesicles ($N=200$), we indeed find a crossover near $c^* \approx 1/(1.6L)$; however, for smaller vesicles ($N=100,\,30$) the transition shifts to much lower curvatures (Supporting Fig.~{S6}), indicating that the curvature-threshold mechanism is not sufficient to fully account for the anchoring behavior in 3D deformable confinement.

We therefore quantify curvature along individual rods by computing $c_{\parallel}(s)$, the membrane curvature projected along the rod axis, measured from the rod midpoint ($s=0$) to its endpoint ($s=L/2$) (Methods and Supporting Fig.~{S5}). In linear vesicles, rods located along the vesicle base exhibit $c_\parallel(s)\approx 0$, whereas rods near the vesicle tips show a sharp rise in $c_\parallel(s)$ as $s \to L/2$ (Fig.~\ref{fig:curv_corr}e), consistent with the observed transition from planar to homeotropic anchoring (Fig.~\ref{fig:curv_corr}a).

Spherical and plate-like vesicles exhibit a qualitatively different curvature pattern (Fig.~\ref{fig:curv_corr}b,c). Rather than displaying two dominant curvature maxima, they develop multiple alternating regions of high and low curvature, giving rise to faceted or polygonal morphologies (Fig.~\ref{fig:curv_corr}d). These curvature modulations correlate strongly with rod position: regions of high curvature coincide with rod endpoints, while flatter regions align with rod midpoints. Consequently, the corresponding $c_{\parallel}(s)$ profiles increase monotonically from the rod midpoint to its endpoint (Fig.~\ref{fig:curv_corr}e). This correlation vanishes in isotropically packed vesicles with comparable $\nu$ (Supporting Fig.~{S7}), confirming that the observed curvature patterns are induced by smectic ordering. 

Together, these results explain why plate-like vesicles incur higher bending energies than linear ones. High-curvature regions are consistently localized at rod endpoints (Fig.~\ref{fig:curv_corr}e). Hence, the intrinsic curvature profile of linear vesicles naturally accommodates multiple smectic layers aligned along the vesicle's long axis. In contrast, in spherical and plate-like vesicles the orientation of smectic layers is not congruent with the global vesicle geometry, necessitating multiple localized regions of high curvature to accommodate rod endpoints, which in turn produce faceted morphologies and elevated bending energies. This mechanism accounts for both the higher energetic cost and the lower abundance of plate-like vesicles observed in Fig.~\ref{fig:bending_energy}. For larger system sizes, the curvature correlation remains similar for linear vesicles but weakens for plate-like morphologies, which display reduced faceting and smoother contours (Supporting Fig.~S8). This trend is consistent with the expectation that curvature-induced boundary effects become less important as vesicle size increases toward the bulk limit. For \textit{fd}-virus, the extrapolation length, characterizing the balance between elastic and boundary effects~\cite{Ribas1995}, is on the order of the rod length~\cite{Lewis2014}, suggesting that curvature correlations should become negligible only for much larger systems.

\subsection*{Controlling rod packing through vesicle shape}

\begin{figure}
	\centering
	\includegraphics[width=0.9\linewidth]{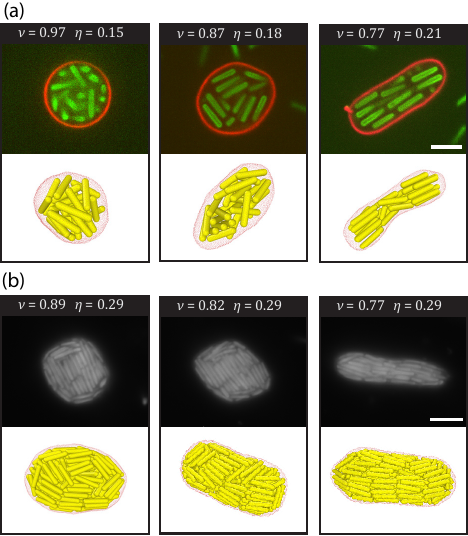}
	\caption{Rod packing under controlled changes in vesicle volume and membrane area. From left to right, panels show the vesicle's configuration as the internal volume decreases (a) and membrane area increases (b). The top row shows experimental snapshots, while the bottom row illustrates simulated vesicles at comparable reduced volume $\nu$ and packing fraction $\eta$ (within 15\%). For the membrane-area modulation experiments, $\nu$ and $\eta$ were estimated from 2D cross-sections (see Supporting Fig.~{S9}). The values reported above each snapshot correspond to those used in the simulations. Scale bars: \qty{5}{\um}.}
	\label{fig:VolumeChange}
\end{figure}

Finally, we demonstrate the versatility of GUVs as soft, tunable containers. The phase diagram in Fig.~\ref{fig:phase} indicates that the packing order of rods can be controlled by adjusting the reduced volume $\nu$ of the vesicle, and thus its shape. To test this prediction, we subjected initially spherical vesicles ($\nu \approx 1$) with isotropically packed rods to gentle osmotic deflation by evaporating the surrounding medium. As the external sugar concentration increased, vesicles deflated at approximately constant membrane area, leading to a decrease in $\nu$, see red-white dashed line in Fig.~\ref{fig:phase}. Simultaneously, solvent efflux decreased the internal volume and increased the packing fraction $\eta$. This combined effect induced a transition from isotropic to nematic and eventually smectic rod ordering, accompanied by a morphological transformation of the vesicles from spherical to linear. Importantly, this process was fully reversible: reinflating the vesicles by slowly diluting the outer medium restored their original shapes and packing states. Representative snapshots of such a transition are shown in Fig.~\ref{fig:VolumeChange}a and Supporting Movie~{S3}.

An alternative route to control rod packing within the $\eta$–$\nu$ phase space is to alter the membrane area while keeping the vesicle volume approximately constant, see vertical red-white dashed line in Fig.~\ref{fig:phase}. Although experimentally more challenging, this approach isolates the effect of vesicle shape $\nu$, since the packing fraction $\eta$ remains fixed. We realized this mechanism by exploiting membrane oxidation induced by fluorescently labeled, polydisperse SU-8 rods encapsulated in the vesicle (see Methods). Upon illumination with 560 nm light, reactive oxygen species generated by the fluorescent dye incorporated in the rods increased the vesicle membrane area by approximately $10\%$, leading to a rapid decrease in $\nu$~\cite{sankhagowit2014dynamics}. Because of the fast kinetics of this transition, full confocal $z$-stacks could not be acquired in real time; consequently, $\nu$ and $\eta$ were estimated from the 2D cross-sections (see Supporting Fig.~{S9}), with simulations ($N=200$) confirming the accuracy of these estimates.

During the membrane-area increase, the initially plate-like vesicle with planar rod anchoring transformed into a linear shape exhibiting nematic order, see Fig.~\ref{fig:VolumeChange}b and Supporting Movies~{S4} and {S5}. In the resulting linear vesicle, rods align parallel to the membrane along the vesicle base and reorient homeotropically at the tips, consistent with the behavior observed for silica rods, despite differences in aspect ratio and polydispersity. Simulations reproduce the same curvature-driven realignment: although the initial vesicle is more spherical due to the absence of gravity in the simulation; the decrease in $\nu$ at fixed $\eta$ drives a transition toward a linear morphology with nematic order. This agreement between experiment and simulation confirms that reducing $\nu$ while maintaining constant packing fraction $\eta$ is sufficient to reorganize the rod packing within the vesicle.

\section*{Conclusion}
To summarize, our study demonstrates that the ordering of rod-like particles within soft, deformable containers gives rise to a competition between anisotropic interactions, geometric confinement, and boundary conditions. This interplay results in a bidirectional coupling between internal rod organization and container morphology. By combining experiments and simulations of colloidal rods encapsulated within lipid vesicles, we show that elongated vesicle shapes promote nematic alignment, effectively shifting the isotropic–nematic transition to lower packing fractions. Conversely, at high densities, frustration of smectic order generates mechanical stresses that drive pronounced vesicle deformations, giving rise to plate-like morphologies. These plate-like vesicles exhibit a significantly higher bending energy than their linear counterparts, illustrating how internal ordering can deform vesicles away from their equilibrium shapes. Finally, we demonstrate the versatility of giant unilamellar vesicles (GUVs) as adaptive soft containers by inducing reversible transitions between isotropic, nematic, and smectic phases, as well as between elongated and plate-like morphologies, through controlled modulation of internal volume or membrane area.

In contrast to previous studies of rigid two-dimensional elliptical hard containers~\cite{Jull2024}, where anchoring conditions could be predicted from the local wall curvature, exhibiting homeotropic anchoring above a critical curvature threshold, we find a fundamentally different mechanism in deformable vesicles. Here, the membrane adapts its curvature to the internal rod organization, maintaining planar anchoring along the rod bodies and curving predominantly near the rod endpoints. In linear vesicles, which exhibit two highly curved tips, this adaptive curvature naturally couples to the rod organization, aligning the director field along the long axis of the vesicle. For spherical and plate-like vesicles, however, such global alignment is not possible. Instead, the system accommodates the internal order through localized deformations, resulting in faceted vesicle geometries characterized by regions of high curvature near rod endpoints and extended flat areas along the rod bodies. 

Our findings reveal the presence of a bidirectional coupling between container shape and rod organization, which is relevant to a range of biological processes, including cell motility, division, and engulfment, in which cytoskeletal filaments are confined by lipid membranes~\cite{fletcher2010cell,pollard2009actin}. While living cells involve additional complex biochemical regulation and specific interactions, our minimal model decouples the physical consequences of flexible confinement and particle anisotropy. Our results demonstrate that, even in the absence of molecular complexity, strong coupling between internal organization and boundary shape naturally emerges. This highlights the role of generic physical interactions in driving organization and deformation, and provides a foundation for understanding more complex cellular and synthetic systems.

Finally, our results could open up the possibility of self-assembly of smectic structures with a new class of materials. Smectic structures are typically self-assembled by droplet evaporation~\cite{Querner2008, Zanella2011}, which constrains the type of materials and the shape of the final structure. By contrast, the use of flexible containers enables smectic self-assembly without complete solvent evaporation, while simultaneously allowing access to a wider range of final morphologies. These include not only spherical structures, similar to those produced by droplet evaporation, but also linear and plate-like configurations. Moreover, our results shed light on the emergence of smectic order under isotropic pressure, complementing previous studies of smectic cluster growth~\cite{Cuetos2010}.

\section*{Acknowledgments}
The authors thank Fabrizio Camerin and Susana Marín-Aguilar for useful discussions. S.v.d.H and H.R.V. thank Frieder Mugele and Mireille Claessens for kindly providing access to confocal and fluorescence microscopes. H.R.V. acknowledges funding from the Sector Plan from The Netherlands Organization for Scientific Research (NWO, Dutch Science Foundation), and partial support from the European Research Council (ERC) under the European Union’s Horizon Europe research and innovation programme (Grant agreement No. ERC-2024-CoG 101171050 – SynthAct3D). A.F.V.M and M.D. acknowledge funding from the European Research Council (ERC) under the European Union’s Horizon 2020 research and innovation program (Grant Agreement No. ERC-2019-ADG 884902 SoftML) and partial support from the European Union’s Horizon Europe research and innovation program under the grant agreement number 101203506, Marie Sklodowska-Curie Action Postdoctoral Fellowship, project IonFlowElast.

\bibliography{biblio}

\section*{Materials and Methods}

\subsection*{Experimental}

\subsubsection*{Materials}

All chemicals were used as received, unless specified otherwise. 1-pentanol (99\%), PVP (M\textsubscript{n} = \qty{40000}{\g/\mol}), absolute ethanol, sodium citrate dihydrate (99\%), ammonia (28\% in water), TEOS (99\%), fluorescein isothiocyanate (FITC) (isomer I, 90\%), APS (98\%), chloroform ($\geq$99.5\%), heavy mineral oil, glucose, sucrose, cholesterol, Rhodamine B isothiocyanate (RITC), and PEG-PPG-PEG Pluronic F-108 were purchased from Sigma Aldrich. The lipids, 1,2-dioleoyl-sn-glycero-3-phosphocholine (DOPC) and fluorescent 1,2-dioleoyl-sn-glycero-3-phosphoethanolamine-N-(lissamine rhodamine B sulfonyl) (ammonium salt) (Liss Rhod PE) in chloroform were obtained from Avanti Polar Lipids (Alabaster, AL). SU-8 50 photoresist was procured from Microchem. All aqueous solutions were prepared using ultrapure Milli-Q\textsuperscript{\textregistered} water. Sugar solutions were filtered through a \qty{0.2}{\um} cellulose filter (VWR). 8-well chamber slides ($\mu$-Slide 8 Well Glass Bottom) were obtained from Ibidi.

Lipid stock solutions were prepared by dissolving or diluting lipids in chloroform to the following concentrations: \qty{12}{\mg/\ml} for DOPC, \qty{0.2}{\mg/\ml} for Liss Rhod PE, and \qty{12}{\mg/\ml} for cholesterol. Stock solutions were stored at \qty{-20}{\degreeCelsius} until use.

\subsubsection*{Synthesis of silica rods}

Fluorescent FITC-coated silica rods were prepared using the method from Ref.~\cite{kuijk2011synthesis}, decreasing the batch size by a factor 10. In brief, rods were grown by adding \qty{30}{\ml} of 1-pentanol, in which 3 g of PVP was dissolved, to a \qty{50}{\ml} centrifuge tube. Then, to this tube, \qty{3}{\ml} absolute ethanol, \qty{0.84}{\ml} ultrapure water, and \qty{0.2}{\ml} of \qty{0.18}{\Molar} sodium citrate dihydrate in water was added. The flask was shaken to mix the contents, followed by the addition of \qty{0.675}{\ml} ammonia. The flask was shaken again, and \qty{0.3}{\ml} TEOS was added, followed by a final shaking. The bottle was then left to allow the reaction to proceed overnight. To elongate the rods, after \qty{20}{\hour}, \qty{150}{\ul} TEOS was added, and another \qty{200}{\ul} TEOS was added \qty{4}{\hour} later. The reaction was stopped after another \qty{4}{\hour}, resulting in a rod length of $\sim$\qty{4.5}{\um}.

The rods were washed with ethanol, water, and again ethanol following the steps described in~\cite{kuijk2011synthesis}. Then, a thin layer of fluorescent silica was grown around the rods by dispersing the rods in \qty{10}{\ml} ethanol. Under magnetic stirring, \qty{1.2}{\ml} ammonia, \qty{1}{\ml} water, \qty{0.5}{\ml} ethanol containing FITC and APS, and \qty{0.1}{\ml} TEOS were added. The FITC-APS solution was prepared by dissolving \qty{25}{\mg} FITC and \qty{35}{\ul} APS in \qty{5}{\ml} ethanol and letting this react overnight. The rod mixture was left to react for several hours, checking small aliquots of the sample intermittently under the microscope to see the fluorescence intensity. Once the fluorescence was strong enough, the reaction was stopped by washing with ethanol.

Finally, to increase the stability of the rods and minimize their interaction with the GUV membrane, as well as increase the distance between the fluorescent signal obtained from neighbouring rods, a final shell of non-fluorescent silica was grown around the rods following the same procedure, except that no FITC solution was added. This final coating was repeated twice to achieve sufficient spacing between the rods. 

Then, the sample was washed with ethanol and water, and cleaned to increase monodispersity, removing small rods by centrifuging (Eppendorf\textsuperscript{\textregistered} Centrifuge 5425) at 120 $g$ and removing clusters by centrifuging at 50 $g$ for \qty{2}{\min} and collecting the supernatant, followed by concentrating the sample. The main population of rods had an average length $L = \qty{4.5}{\um}$ and diameter $D = \qty{1}{\um}$ (see Supporting Fig.~{S1}), corresponding to an aspect ratio of $L/D = 4.5$, as measured in scanning electron microscopy (SEM, JEOL JSM-6010A SEM at 5 kV).

\subsubsection*{Synthesis of SU-8 rods}

RITC-containing microrods were fabricated following an adapted protocol for SU-8 rods from~\cite{fernandez2019synthesis}, based on earlier work by Alargova \textit{et al.}~\cite{alargova2004scalable,alargova2006formation}. In a typical synthesis, \qty{110}{\mL} of glycerol was added to a \qty{250}{\mL} tall beaker. A mixer (Ika Werk RW-20, Janke\&Kunkel) equipped with a \qty{4}{\cm} impeller was positioned \qty{2}{\cm} above the bottom of the beaker and set to 2000 rpm. Next, approximately \qty{0.1}{\g} of SU-8 50, containing 0.04 wt.-\% RITC, was added to the glycerol between the impeller and the beaker wall. Mixing was continued for \qty{10}{\min}, after which the beaker was covered with aluminum foil to protect it from light and placed in a sonication bath (M2800H-E, Branson) for \qty{2}{\hour} to fragment the rods. The sonicated glycerol solution was then exposed to UV light (XL-1500 V crosslinker; 6 × \qty{15}{\watt}, \qty{365}{\nm}, Spectrolinker) for \qty{15}{\min} to crosslink the polymer rods.

In order to replace the glycerol with water, rods were washed through repeated cycles of centrifugation and redispersion in Milli-Q\textsuperscript{\textregistered} water containing 0.5 wt.-\% Pluronic F108. Centrifugation was performed at 3000 $g$ for \qty{30}{\min}, and redispersion was achieved by vortexing and sonication for \qty{10}{\min}. To remove large or irregular rods, the sample was allowed to sediment for \qty{1}{\hour}, and the precipitate was discarded. This was followed by four cycles of centrifugation at 600 $g$ for \qty{10}{\min}, in which the supernatant was collected and the precipitate redispersed for the next cycle. After the final cycle, the precipitate was discarded, and the accumulated supernatant was centrifuged at 3000 $g$ for \qty{30}{\min}. The resulting precipitate was redispersed in Milli-Q\textsuperscript{\textregistered} water containing 0.5 wt.-\% Pluronic F108 to a desired rod concentration.

\subsubsection*{Preparation of GUVs}

Giant unilamellar vesicles (GUVs) were prepared using a modified droplet transfer approach adapted from Ref.~\citealp{vutukuri2020active} and~\citealp{vanderHam2024shape}. To prepare the lipid-in-oil solution (LOS), \qty{100}{\ul} of DOPC (\qty{12}{\mg/\ml}) and \qty{10}{\ul} of Liss Rhod PE (\qty{0.2}{\mg/\ml}) stock solutions were combined in a \qty{20}{\ml} glass vial. The chloroform solvent was removed by gently flowing N\textsubscript{2} over the mixture while rotating the vial to form a uniform lipid film at the bottom. The film was then placed under vacuum in a desiccator for 1–2 \unit{\hour} to remove any remaining solvent. Subsequently, \qty{3}{\g} of heavy mineral oil was added, giving a final lipid concentration of \qty{425}{\micro\Molar} with \qty{0.1}{\mol\percent} Liss Rhod PE. The mixture was sonicated at \qty{40}{\degreeCelsius} for \qty{1}{\hour} and stored overnight in the dark at room temperature to ensure complete lipid dissolution.

For the osmotic inflation experiments, cholesterol was included in the GUV membrane to suppress spontaneous curvature arising from leaflet lipid-number asymmetry~\cite{bhatia2020simple}. The same protocol as described above was followed with two modifications. The lipid composition was adjusted to \qty{70}{\ul} of DOPC (\qty{12}{\mg/\ml}), \qty{60}{\ul} of Liss Rhod PE (\qty{0.2}{\mg/\ml}), and \qty{100}{\ul} of cholesterol (\qty{3}{\mg/\ml}). At a cholesterol incorporation efficiency of $\sim20\%$~\cite{weakly2024several}, this results in $\sim \qty{15}{\mol\percent}$ cholesterol in the membrane.

To form the GUVs, \qty{200}{\ul} of LOS was carefully layered over \qty{500}{\ul} of outer solution (\qty{100}{\milli\Molar} glucose) in a \qty{2}{\ml} Eppendorf\textsuperscript{\textregistered} tube. In a separate \qty{2}{\ml} tube, \qty{200}{\ul} of LOS was mixed with \qty{15}{\ul} of inner solution, which consisted of \qty{100}{\milli\Molar} sucrose containing silica rods. For the area change experiments, the silica rods were replaced by the SU-8 rods. The mixture was emulsified by mechanically agitating the tube 25 times on a tube rack, producing water-in-oil droplets~\cite{moga2019optimization}. Then, \qty{120}{\ul} of the resulting emulsion was gently layered onto the water-oil column in the first tube and centrifuged immediately at \qty{200}{\g} for \qty{2}{\min}. After centrifugation, the top oil layer was carefully removed, and the GUVs were allowed to settle for \qtyrange{30}{60}{\min} before use in experiments.

\subsubsection*{Sample preparation}

To study rod packing in GUVs, \qty{20}{\ul} of the GUV solution, taken from the bottom of the tube, was gently added to one of the wells of the 8-well chamber slide, which was prefilled with \qty{280}{\ul} of \qty{100}{\milli\Molar} glucose. For the osmotic inflation experiments, the well was filled with a 2:1 mixture of \qty{100}{\milli\Molar} sucrose and glucose to minimize the sugar asymmetry. Typically, the well was left open over the course of several hours to evaporate the outer solution, thus increasing the sugar concentration and deflating the GUVs. 

\subsubsection*{Confocal microscopy imaging}
Microscopy measurements were performed on a confocal laser scanning microscope (Nikon Eclipse Ti-U inverted microscope with VTinfinity3 CLSM module, Visitech) equipped with a Hamamatsu ORCA-Flash4.0 CMOS camera and a 100×, 1.49 NA oil-immersion objective. GUVs labeled with Liss Rhod PE were excited using a \qty{561}{\nm} laser. Time-lapse imaging was carried out at 5-10 frames per second, while $z$-stacks were acquired with a step size of \qty{0.10}{\um} or \qty{0.25}{\um} and an exposure time of \qty{50}{\ms}, corresponding to a scan rate of roughly 8 slices per second.

\subsubsection*{Reduced volume and packing fraction}
The vesicle's reduced volume $\nu$ was obtained by measuring the vesicle volume $V_\mathrm{ves}$ and membrane area $A_\mathrm{ves}$ and inserting these values into Eq.~\ref{eq:reduced_volume}. In experiments, $V_\mathrm{ves}$ and $A_\mathrm{ves}$ were extracted from confocal $z$-stacks using the LimeSeg plugin in Fiji (ImageJ)~\cite{schindelin2012fiji,machado2019limeseg}.

Because imaging was performed in an aqueous medium with an oil-immersion objective, the $z$-spacing of the stacks was corrected by a factor of 0.83 to account for spherical aberration. This factor was calculated from the numerical aperture (1.49), the refractive index of the imaging medium (1.33), and the refractive index of the immersion oil (1.52)~\cite{diel2020tutorial,van2025shallow}.

The packing fraction of the rods, $\eta$, was estimated by counting the number of encapsulated rods $N$, multiplying by the average rod volume $V_\mathrm{rod}$, and dividing by $V_\mathrm{ves}$ (see Eq.~\ref{eq:packing_fraction}). All rods were counted, including those shorter than the dominant length population. Accordingly, $V_\mathrm{rod}$ was calculated using the average length, $\langle L \rangle = \qty{3.99}{\um}$, and diameter, $\langle D \rangle = \qty{1.01}{\um}$, of the full population (see Supporting Fig.~{S1}). Each rod was approximated as a cylinder with one spherical cap and one flat end (bullet-like geometry)~\cite{kuijk2011synthesis}, giving $V_\mathrm{rod} = \pi (R_\mathrm{rod})^2 (L_\mathrm{rod}-R_\mathrm{rod}) + \frac{2}{3}\pi R_\mathrm{rod}^3 \approx \qty{3.1}{\um\cubed}$, where $L_\mathrm{rod} = \qty{3.99}{\um}$ is the full length of the rod, and $2R_\mathrm{rod} = \qty{1.01}{\um}$ is the diameter of the cylindrical section. 

\subsubsection*{Vesicle membrane segmentation}
The vesicle membrane was segmented from confocal fluorescence $z$-stacks and reconstructed as a three-dimensional surface mesh. Image stacks were imported into FIJI (ImageJ) and segmented using the LimeSeg plugin~\cite{machado2019limeseg,schindelin2012fiji}. The resulting surface meshes were exported as point clouds and further processed using a custom Python analysis pipeline.

From each point cloud, a triangular surface mesh was generated using Poisson Surface Reconstruction~\cite{kazhdan2006poisonsurface} (Open3D function \texttt{create\_from\_point\_cloud\_poisson}) at high resolution. To reduce computational cost while preserving geometric accuracy, the mesh was subsequently downsampled using quadric error metric surface simplification~\cite{garland1997surfacesimplify} (\texttt{simplify\_quadric\_decimation}) to a target resolution corresponding to an effective vertex radius $r_\mathrm{vertex} = \qty{0.2}{\um}$, defined as $r_\mathrm{vertex} = (A_\mathrm{vertex}/\pi)^{1/2}$,
where $A_\mathrm{vertex}$ is the average surface area associated with each vertex.

Local membrane curvature was computed on the resulting meshes using the \texttt{principal\_curvature} function from the libigl library~\cite{libigl}. This method estimates the two principal curvatures $c_1$ and $c_2$ and their corresponding directions $\mathbf{v}_1$ and $\mathbf{v}_2$ at each vertex via quadric surface fitting over the local mesh neighborhood.

Shape classification of the vesicle mesh was performed as described in the \textit{Vesicle shape classification} section below.

\subsubsection*{Rod segmentation}
Rod centroids, orientations, and lengths were obtained by three-dimensional segmentation of the fluorescence $z$-stack containing the rod signal. Due to the limited axial resolution of the confocal microscope, direct three-dimensional thresholding was not feasible, as it led to merging of rods along the $z$-direction. Instead, segmentation was performed slice-by-slice, exploiting the higher lateral resolution in the $xy$-plane.

Each individual $xy$ slice was segmented using a local Otsu threshold with a radius of 15 pixels in FIJI ImageJ~\cite{schindelin2012fiji}, resulting in a binarized image stack. This binarization was refined using successive morphological opening and closing operations applied independently to each slice.

Connected components in each slice were subsequently linked across consecutive slices using a custom Python script. Components were matched based on similarity in centroid position, major and minor axis lengths, and in-plane orientation. A similarity threshold was imposed to distinguish individual rods along the $z$-direction, while components that were either too small (length $\leq \qty{2}{\um}$) or present in too few $z$-slices (e.g., excessively flat objects) were filtered out. Finally, manual corrections to the binarized stack were applied to optimize rod segmentation, with detected rod centerlines overlaid on the original image stack for validation. The resulting three-dimensional connected components were then analyzed using principal component analysis (PCA) to determine their centroid, the unit vector along the major axis $\mathbf{u}_i$, and the rod length $L$.

Classification of the rod organization as isotropic, nematic, or smectic was performed as described in the \textit{Nematic and smectic order} section below.

\subsubsection*{Curvature correlation}
To determine the correlation $c_\parallel(s)$ between membrane curvature and position along a rod’s long axis, the segmented vesicle membrane and rod geometries were combined as follows. Each rod was divided into 20 equally spaced segments along its length. The distance of a segment from the rod midpoint is denoted by $s$ and treated as positive due to rod symmetry.

For each rod segment at position $s$, the nearest vertex on the vesicle surface mesh was identified. If this vertex lay within a cutoff distance of $1.5D$ from the rod segment, the local membrane curvature was assigned to that segment. At each mesh vertex, the two principal curvatures $c_1$ and $c_2$ with corresponding directions $\mathbf{v}_1$ and $\mathbf{v}_2$ span the local tangent plane. The rod orientation vector $\mathbf{u}$ was projected onto this plane to obtain $\mathbf{u}_\mathrm{proj}$, and the curvature along the rod direction was computed as $c_\parallel = \left( \theta_1 c_2 + \theta_2 c_1 \right) / \left(\theta_1 + \theta_2 \right)$,
where $\theta_1$ and $\theta_2$ are the angles between $\mathbf{u}_\mathrm{proj}$ and $\mathbf{v}_1$ and $\mathbf{v}_2$, respectively. A schematic illustration of this construction is shown in Supporting Fig.~{S5}.

This procedure was repeated for all segments and all rods within a vesicle, and the resulting values were averaged as a function of $s$ to obtain the curvature–position correlation $c_\parallel(s)$.

\subsubsection*{Osmotic inflation of GUVs}
After deflating the GUVs, linear smectic configurations were selected that did not exhibit the formation of tubes/buds, indicative of near zero spontaneous curvature. The lid of the well was then replaced by a customized version in which a tube, connected to a syringe pump, was added. This tube allowed for controlled and gentle inflow of Milli-Q\textsuperscript{\textregistered} water, which caused a decrease in the sugar concentration, leading to an inflation of the GUVs. The syringe pump was set to \qty{10}{\ul\per\min}.

\subsection*{Numerical simulations}

\subsubsection*{Interaction potentials}

The particles interact via a soft repulsive Weeks-Chandler-Andersen (WCA) potential:
\begin{equation}
	U_{WCA}(r) = 
	\begin{cases}
		4\epsilon \left[\left(\frac{\sigma_T}{r}\right)^{12}-\left(\frac{\sigma_T}{r}\right)^6\right] + \epsilon & \mathrm{if}~r\leq 2^{1/6}\sigma_T\\
		0 & \mathrm{otherwise,}
	\end{cases}
\end{equation}
where $\epsilon$ sets the interaction strength, $r$ is the center-to-center distance between two particles, and $\sigma_T$ is the sum of the radii of the interacting particle pair. The particle diameters are $\sigma$ for the spheres composing the rods and $\sigma_v=\sigma/4$ for both the vesicle particles and the explicit solvent.

Vesicle-vesicle interactions are modeled using a meshless membrane model. Full details of the model are provided in Refs.~\citealp{Yuan2010,Fu2017}. The membrane is composed of spheres with diameter of $\sigma/4$. The model combines a short-range repulsive interaction with a long-range attractive potential, characterized by the following parameters: $r_\mathrm{min}$, the distance at which the potential is minimized; $r_c$, the cutoff distance; $\zeta$ a dimensionless correction factor for the attractive interaction; $\mu$, a dimensionless bending rigidity parameter; and $\sin \theta_0$ sets the spontaneous curvature. The parameter values used in this study are listed in Table~\ref{tab:sim_parameters}. We set the membrane parameters such that the vesicle does not yield spontaneous curvature and has a bending rigidity of $20 k_\mathrm{B} T$, comparable to that of the experimental DOPC vesicles~\cite{faizi2020fluctuation}.

\begin{table}
	\centering
	\caption{Parameters of the interaction potentials used in the simulations}
	\begin{tabular*}{\hsize}{@{\extracolsep{\fill}}lr|lr}
		\textbf{Parameter} &    \textbf{Value} & \textbf{Parameter} & \textbf{Value} \cr \hline
		$\epsilon$         &        $4.35k_\mathrm{B}T$ & $\zeta$            &          $4.0$ \cr
		$\sigma_v$         &      $0.25\sigma$ & $\mu$              &          $3.0$ \cr
		$r_\mathrm{min}$   & $2^{1/6}\sigma_v$ & $\sin \theta_0$    &            $0$ \cr
		$r_c$              &     $2.6\sigma_v$ &                    &                \cr \hline
	\end{tabular*}
	\label{tab:sim_parameters}
\end{table}

\subsubsection*{Simulation protocol}
Simulations are initialized with a perfectly spherical vesicle of diameter $D_\mathrm{v}$, containing an ordered assembly of rods, and surrounded by explicit solvent particles arranged on a regular grid. The vesicle particles are initially distributed uniformly over the spherical surface using the following azimuthal ($\varphi$) and polar ($\theta$) spacings
\begin{align}
	d\varphi &= \frac{\pi}{2\arcsin(0.97/\sigma_v)} \mathrm{ ,} \\
	d\theta  &= \frac{\pi \sin \varphi}{\arcsin(0.97/\sigma_v)} \mathrm{ .}
\end{align}
Rods are packed by stacking multiple $n\times n$ rod arrays, where $n$ is the number of spheres in each rod, until the desired number of rods is reached. The rod orientations are chosen to minimize the radius of gyration of the rod assembly, enabling placement within a compact spherical vesicle and allowing for high packing fractions, see Fig.~\ref{fig:setup}c. The explicit solvent is placed outside the vesicle on a regular cubic lattice of side length $4D_\mathrm{v}$, with a lattice spacing of $2.7\sigma_v$ to prevent overlap between solvent particles. As a result, the simulation box always fully encloses the vesicle.

The system is equilibrated by slowly compressing the explicit solvent, from zero pressure up to the desired pressure, using the $NPT$ ensemble in LAMMPS. The vesicle and rods are evolved in the $NVT$ ensemble. The compression protocol spans $2\times10^6$ iterations, thus at the highest compression rate, this corresponds to an average pressure increase of $\beta\Delta P \sigma^3/\Delta t\approx10^{-5}$ per iteration. Following compression, time averages are collected every $5\times10^4$ iterations over a total of $10^6$ iterations. For the largest vesicles and highest pressures, the meshless membrane occasionally ruptures, resulting in the formation of two vesicles, as previously reported in Ref.~\cite{Fu2017}. Such trajectories are identified and excluded from the analysis using a clustering algorithm.

The rods are constructed from spheres that are constrained to move as rigid bodies. To simulate vesicles containing free spheres, the rod spheres are instead allowed to move independently, while all other aspects of the initial configuration and simulation protocol are kept identical.

In simulations of the vesicle deflation, after equilibration, the system is slowly compressed from $\beta P\sigma^3 = 0.83$ to $14$ over $10^6$ iterations, resulting in a compression rate per iteration of $\beta\Delta P \sigma^3/ \Delta t \approx 10^{-5}$. In simulations of the vesicle area expansion, the system is first equilibrated, after which the vesicle particle diameter is linearly increased up to $1.025\sigma_v$ over $10^6$ iterations. This protocol results in a surface area increase of $A/A_0\approx1.05^2=110\%$.

\subsubsection*{Reduced volume and packing fraction}

The vesicle's reduced volume, $\nu$, and the rod packing fraction, $\eta$, were calculated by first constructing the vesicle surface mesh from the positions of the membrane particles (as described in the Experimental section). The resulting vesicle volume, $V_\mathrm{ves}$, and membrane area, $A_\mathrm{ves}$, were then inserted into Eq.~\ref{eq:reduced_volume} and Eq.~\ref{eq:packing_fraction}, respectively, to compute $\nu$ and $\eta$.

\subsubsection*{Vesicle shape classification}

To classify vesicle shapes, we use the asphericity $b$ and acylindricity $c$ defined as 
\begin{align}
	b &= \frac{3}{2}\lambda_z^2 - \frac{R_g^2}{2} \\
	c &= \lambda_y^2 - \lambda_x^2
\end{align}
where $\lambda_x^2 \leq \lambda_y^2 \leq \lambda_z^2$ are the eigenvalues of the radius of gyration tensor, $S_{mn}=1/N\sum_{i=1}^N r_m^{(i)} r_n^{(i)}$, where $r_m^{(i)}$ is the $m$ component of the position of membrane particle $i$, and $R_g=\sqrt{\lambda_x^2+\lambda_y^2+\lambda_z^2}$ is the radius of gyration. By definition, $b=0$ corresponds to perfect spherical symmetry, while $c=0$ corresponds to perfect cylindrical symmetry; larger values indicate more anisotropic vesicles. We classify vesicles as spherical if $b/R_g^2\lesssim0.25$, linear if $c/R_g^2\lesssim0.11$, and plate-like if neither condition is fulfilled. These thresholds were chosen based on visual inspection of representative vesicles, and the exact boundaries between each shape region are not sharply defined.

\subsubsection*{Nematic and smectic order}

Due to the low number of rods, conventional nematic and smectic order parameters are not suitable to classify rod ordering. Instead, we classify the nematic order based on the local and smectic order parameters. The local nematic order of each rod $i$ is~\cite{Cuetos2007}
\begin{equation}
	S_i = \frac{1}{n_i} \sum_{j=1}^{n_i}\left[\frac{3}{2}\left(\mathbf{u}_i \cdot \mathbf{u}_j\right)^2 - \frac{1}{2}\right] \\,
\end{equation}
where $\mathbf{u}_i$ is the unit vector along the major axis of rod $i$, and $n_i$ is the number of neighboring rods of particle $i$, defined as those with a surface-to-surface distance $\rho_{ij} \leq \sigma$. The local smectic order parameter of each rod is determined using~\cite{Bakker2016}
\begin{equation}
	\tau_i = S_i \left[1 - \frac{1}{m_i}\sum_
	{j=1}^{m_i}\frac{\mathbf{r}_{ij}\cdot \mathbf{u}_i}{r_\mathrm{cut}}\right] \\,
\end{equation}
where $r_\mathrm{cut}=L/2$, and $m_i$ is the number of neighbors of particle $i$ satisfying $r_{ij}<r_\mathrm{cut}$, with $r_{ij}$ the center-to-center distance of the rod pair. The nematic order of each vesicle is determined according to the average local order parameter: isotropic if $\langle S_i \rangle < S_\mathrm{crit}$ and $\langle \tau_i \rangle < \tau_\mathrm{crit}$, nematic if $\langle S_i \rangle > S_\mathrm{crit}$ and $\langle \tau_i \rangle < \tau_\mathrm{crit}$, and smectic if $\langle S_i \rangle > S_\mathrm{crit}$ and $\langle \tau_i \rangle > \tau_\mathrm{crit}$, where the thresholds are $S_\mathrm{crit}=0.4$ and $\tau_\mathrm{crit}=0.375$. For $L/D=8$ we consider $S_\mathrm{crit}=0.5$ and $\tau_\mathrm{crit}=0.6$ due to a sharper increase in the local order parameters. See Fig.~\ref{fig:setup}c for examples of vesicles with different nematic order. These values were determined with visual inspection of the vesicles guided by the inflection point of the local order parameters as a function of the osmotic pressure, see Supporting Fig.~{S10}. We remark that the threshold choice impacts the phase identification.



\subsubsection*{Curvature analysis}

The vesicle mesh, local curvature, and the calculation of $c_\parallel(s)$ were performed using the same workflow described in the Experimental section, with the positions of the membrane particles serving as the input point cloud. The positions and orientations of the rods were taken directly from the simulation output. 

\subsubsection*{Bending energy}

Using the curvature fields obtained from the simulation data, as described above, the mean curvature $M_v$ at each mesh vertex $v$ was computed from the principal curvatures as $M_v = (c_1 + c_2)/2$. The bending energy of the membrane, $E_\mathrm{b}$, was then estimated by summing the squared mean curvature weighted by the corresponding vertex area $A_v$:
\begin{equation}
	E_\mathrm{b} = 2\kappa \int M^2 \, \mathrm{d}A \;\approx\; 2\kappa \sum_v M_v^2 A_v \\,
\end{equation}
where $\kappa$ is the bending rigidity of the membrane.

\end{document}